\documentclass[prd, twocolumn, superscriptaddress, nofootinbib]{revtex4}
\usepackage{epsfig}

\newcommand{\beq}{\begin{equation}}
\newcommand{\eeq}{\end{equation}}
\newcommand{\beqa}{\begin{eqnarray}}
\newcommand{\eeqa}{\end{eqnarray}}
\newcommand{\om}{\Omega_m}
\newcommand{\omc}{\Omega_{rc}}

\begin{document} 

\title{Exploring the Expansion History of the Universe}
\author{Eric V. Linder}
\affiliation{Physics Division, Lawrence Berkeley National Laboratory, 
Berkeley, CA 94720} 

\begin{abstract} 
Exploring the recent expansion history of the universe 
promises insights into the cosmological model, the nature of 
dark energy, and potentially clues to high energy 
physics theories and gravitation.  We examine the extent to 
which precision distance-redshift observations can map out 
the history, including the acceleration-deceleration transition,  
and the components and equations of state of the energy density. 
We consider the ability to distinguish between various 
dynamical scalar field models for the dark energy, as well as 
higher dimension and alternate gravity theories.  Finally, we 
present a new, advantageous parametrization for the study of 
dark energy. 
\end{abstract} 

\maketitle 

\section{Introduction}\label{intro} 

The quest to explore the expansion 
history of the universe has carried cosmology well beyond 
``determining two numbers'' -- the present 
dimensionless density of matter $\Omega_m$ and the present 
deceleration parameter $q_0$ \cite{san61}.  Observations 
have advanced so that now cosmologists 
seek to reconstruct the entire 
function $a(t)$ representing the expansion history of the 
universe.  

A myriad of cosmological observational tests can probe the 
function $a(t)$ more fully, over much of the age of the universe 
(see Sandage \cite{san88}, Linder \cite{lin89,lin97}, Tegmark 
\cite{teg}).  
This paper concentrates on the most 
advanced method, the magnitude-redshift relation of Type Ia supernovae. 

Just as understanding the thermal history of the early universe 
has taught us an enormous amount about both cosmology and 
particle physics (see, e.g., \cite{kt}), the recent expansion history 
of the universe promises similarly fertile ground with the discovery of 
the current acceleration of the expansion of the universe.  This 
involves concepts of the late time role of high energy field theories 
in the form of possible quintessence, scalar-tensor gravitation, higher 
dimension theories, brane worlds, etc.~as well as the fate of the 
universe.

\section{Mapping the Expansion History}\label{sec.map}

Type Ia supernovae, or any standardizable candles (sources with 
known luminosity), are excellently suited to map the expansion history 
$a(t)$ since there exists a direct relation between the observed 
distance-redshift relation $d(z)=(1+z)\eta(z)$ 
and the theoretical $a(t)$.  For a flat universe, assumed throughout, 
\beq 
\eta(z)=\int_{a_e}^1 da/(a^2H)=\int_0^z dz'/H(z')\,,
\label{rh}\eeq 
where $\eta$ is the conformal time, $d\eta=dt/a$, 
the Hubble parameter is 
$H=\dot a/a$ and $a_e=1/(1+z)$ is the scale factor at the time of 
emission, i.e.~when the supernova exploded.  

The Hubble parameter in general relativity comes from 
the Friedmann equation 
\beq
H^2=(8\pi/3)\rho\,,
\label{hrho}\eeq 
and the conservation condition of each component is 
\beq
\dot\rho/\rho=-3H(1+p/\rho)\equiv -3H[1+w(z)],
\label{rhocons}\eeq 
where the energy density is $\rho$, the pressure $p$, 
and the equation of state (EOS) of each component is defined by 
$w=p/\rho$.  Ordinary nonrelativistic matter has $w=0$; 
a cosmological constant has $w=-1$.  We explicitly allow the 
possibility that $w$ evolves.  

The expansion history $a(t)$ is given by 
\beq
t(a)=\int_a^1 da'/(a'H)=\int_0^z dz'/[(1+z')H(z')].
\label{ta}\eeq 

The fly in the ointment is that the measured 
distance $d(z)$ is related to, but 
is not, the desired history relation $a(t)$.  To translate 
$d(z)$ into $a(t)$ directly involves a 
derivative of $d(z)$, so noisy data can introduce difficulties 
\cite{astier,ht}.  Instead one finds $H(z)$ through the cosmology 
parameters $\rho$ and $w(z)$.  

Observational evidence for accelerated 
expansion informs us that there must 
be a component with 
a strongly negative EOS -- ``dark energy'' -- in addition to matter. 
Then combining equations (\ref{rh})-(\ref{rhocons}): 
\beqa
H_0\eta(z)&=&\int_0^z dz'\,\Big[\om(1+z')^3 \label{rwz} \\ 
& &\mbox{}+(1-\om) e^{-3\int_0^{\ln(1+z')} 
d\ln(1+z'') [1+w(z'')]}\Big]^{-1/2}, \nonumber 
\eeqa
where $\om$ is the dimensionless matter density $8\pi\rho_m/(3H_0^2)$ 
and $H_0=H(0)$ is the Hubble constant. 
Eq.~(\ref{ta}) can then be used to obtain $a(t)$. 

To find $w(z)$ one could solve the scalar field 
equation for a particular theory but this does not allow 
a model independent parameter space in which to compare models. 
Instead, for generality of treatment, 
various parametrizations of $w(z)$ are used.  

\subsection{Linear $w(z)$}\label{sec.linpar}

The conventional first order expansion to the EOS, including 
the critical property of time variation in the EOS, 
is $w(z)=w_0+w_1z$.  In this case the 
exponential in (\ref{rwz}) resolves to $(1+z)^{3(1+w_0-w_1)}e^{3w_1z}$. 
However this grows increasingly unsuitable 
at redshifts $z>1$.  Analyzing CMB constraints on the distance to 
the last scattering surface at $z_{lss}=1100$ would be problematic. 

\subsection{A new parametrization of the equation of 
state}\label{sec.newpar}

To extend parametrization of dark energy to redshifts $z>1$, 
I suggest a new model:  
\beqa
w(a)&=&w_0+w_a(1-a) \\ 
&=&w_0+w_a z/(1+z). 
\label{wa}\eeqa 
Here the exponential in (\ref{rwz}) resolves to 
$a^{-3(1+w_0+w_a)} e^{-3w_a(1-a)}$. 

This new parametrization has several 
advantages: 1) a manageable 
2-dimen\-sional phase space, 2) reduction to the old linear redshift 
behavior at low redshift, 3) well behaved, bounded behavior 
for high redshift, 4) high accuracy in reconstructing many 
scalar field equations of state and the resulting distance-redshift 
relations, 5) good sensitivity to observational data, 6) simple 
physical interpretation.  

Beyond the bounded behavior, the new parametrization is 
also more accurate than the old one.  For example, in comparison 
to the exact solution for the supergravity inspired SUGRA model 
\cite{braxm} it is accurate in 
matching $w(z)$ to -2\%, 3\% at $z=0.5$, 1.7 vs.~6\%, -27\% for 
the linear $z$ approximation (the constants $w_1$, $w_a$ 
are here chosen to fit at $z=1$).  Most remarkably, 
it reconstructs the distance-redshift behavior of the SUGRA 
model to 0.2\% over the entire range out to the last scattering 
surface ($z\approx1100$). 

Note that $dw/d\ln(1+z)|_{z=1}=w_a/2$; one might 
consider this quantity a natural measure of time variation (it is 
directly related to the scalar field potential slow roll factor $V'/V$) 
and $z=1$ 
a region where the scalar field is most likely to be evolving as the 
epoch of matter domination changes over to dark energy 
domination.  

The future Supernova/Acceleration Probe (SNAP: \cite{snap}) 
will be able to determine $w_a$ to better than $\pm0.55$ (one 
expects roughly $w_a\approx 2w_1$), with use of a prior on $\om$ of 
0.03, or to better than 0.3 on incorporating data from the Planck CMB 
experiment \cite{planck}. For the advantages of combining supernova 
and CMB data see \cite{fhlt}.  The CMB information can be folded in 
naturally in this parametrization, without imposing artificial 
cutoffs or locally approximating the likelihood surface as required 
in the $w_1$ parametrization. In fact, the new 
parametrization is even more promising since the sensitivity of the 
SNAP determinations increases for $w_0$ more positive than $-1$ 
or for positive $w_a$ (see, e.g., \cite{astier}): the values quoted 
above were for a fiducial cosmological constant model.  For example, 
SUGRA predicts $w_a=0.58$ and SNAP would put error bars of 
$\sigma(w_a)\approx0.25$ on that; this would demonstrate time 
variation of the EOS at the 95\% confidence level.  
Incorporation of a Planck prior can improve this to the $\approx$99\% 
confidence level: $\sigma(dw/d\ln(1+z)|_{z=1})\approx0.1$.

\subsection{Expansion and density histories}\label{sec.plots}

\begin{figure}[!hb]
\begin{center} 
\psfig{file=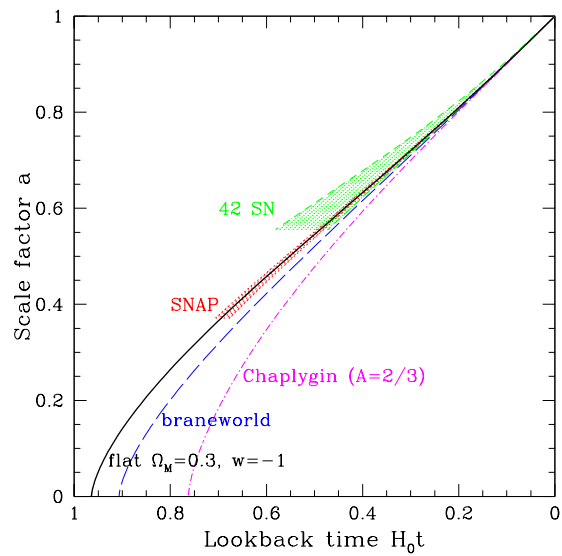,width=3.0in} 
\caption{Mapping the expansion history through the supernova 
magnitude-redshift relation can distinguish the dark 
energy explanation for the accelerating universe from alternate 
theories of gravitation, high energy physics, or higher dimensions. 
All three models take an $\Omega_M=0.3$, flat universe but differ 
on the form of the Friedmann expansion equation. 
} 
\label{atmodsn}
\end{center} 
\end{figure}

Figure~\ref{atmodsn} shows the mapping of the expansion history of 
various models and the constraints that SNAP data would impose. 
The models include both dark energy and alternative explanations 
for the acceleration discussed in Sec.~\ref{sec.bey}.  Regarding the 
data constraints, it is important to note the  
presence of correlation between the parameters $\om$, $w_0$, and 
$w_1$ or $w_a$ (see, e.g., \cite{welal}).  Despite 
SNAP being able to determine each individually to high precision, 
e.g.~$\om$ to 0.03, $w_0$ to 0.05, $w_1$ to 0.3 (each marginalized 
over others), the correlations 
among them relax the tightness of the constraint SNAP would 
place on the expansion history.  This is unavoidable (but see 
Sec.~\ref{sec.conformal}).

\begin{figure}[!hb]
\begin{center} 
\psfig{file=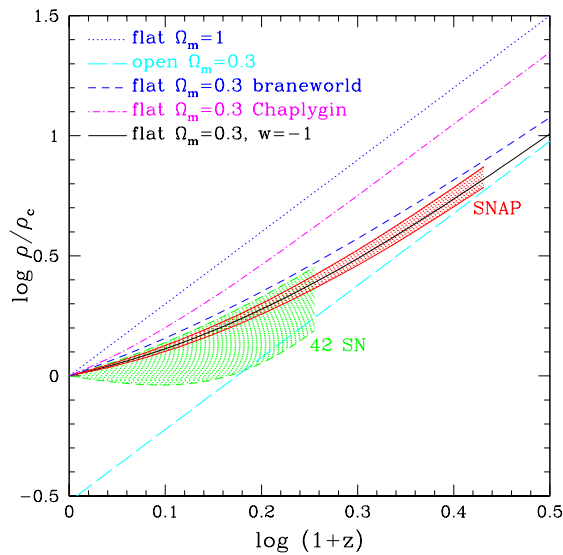,width=3.0in} 
\caption{Mapping the density history with SNAP 
can clearly show both the current accelerating phase and the transition 
to the matter dominated, decelerating epoch.  
At high redshifts all models obey the matter dominated slope 3 law, 
but deviation from this -- the sign of dark energy or alternative 
gravity -- is clearly visible for $z<2$. 
} 
\label{rhosnap}
\end{center} 
\end{figure}

Figure~\ref{rhosnap} gives the equivalent density history. 
Note that the 
matter dominated epoch is characterized by $\rho(z)\sim 
(1+z)^3$ and so has a slope of 3 in this plane.  The deviation 
from this due to the recent epoch of dark energy domination  
can clearly be seen (cf.~Tegmark \cite{teg}). 

\subsection{Conformal time history}\label{sec.conformal}

One method of incorporating the advantages of both 
mapping approaches -- 
the generality of parametrization and the directness of 
reconstruction -- is to 
consider the conformal time $a(\eta)$.  From $d=(1+z)\eta$ 
one sees that one requires no foreknowledge 
or local approximation to obtain the scale factor-conformal 
time relation. 
The error estimation from the observed magnitudes $m$ is simply 
\beq 
{\sigma_\eta\over\eta}=\sigma_m{\ln10\over 5}\approx(1/2)\sigma_m. 
\label{sigeta}\eeq 
A 1\% distance 
measurement error ($\sigma_m=0.02$) given by SNAP's limiting 
systematics becomes a 1\% error in $a(\eta)$.  

Another virtue of the conformal time history is its physical 
interpretation.  Consider the 
logarithmic derivative of the $\eta(a)$ curve.  This is 
\beq 
{d\eta\over d\ln a}=a{dt\over a\,da}=(\dot a)^{-1}.
\eeq 
Two simple physical interpretations apply: 1) this gives the 
proper time evolution of the scale factor, 2) this represents 
the conformal horizon scale $(aH)^{-1}$.  Both are familiar 
from analysis of inflationary cosmology -- they determine 
when the expansion of the universe enters an accelerating  
phase.  

\begin{figure}[!hbt]
\begin{center} 
\psfig{file=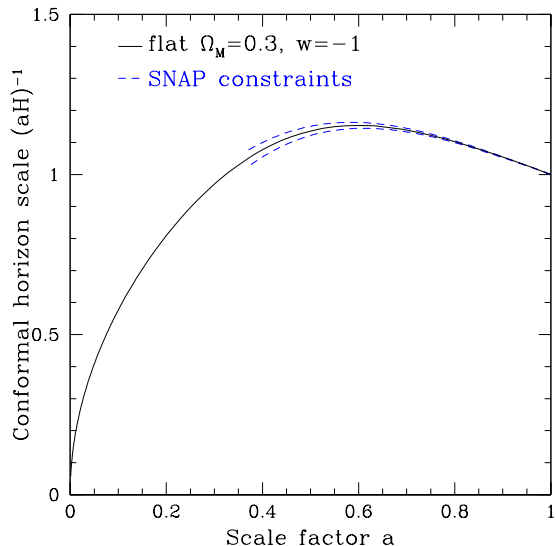,width=3.0in} 
\caption{The 
conformal horizon scale characterizes the dynamics. 
The negative slope part of the curve 
allows comoving wavelengths to expand outside the horizon, or alternately 
represents $aH=\dot a$ increasing, i.e.~$\ddot a>0$ -- the signature of 
inflation or acceleration.  
The dashed blue lines show that SNAP will map the accelerating 
phase, the transition, and into the matter dominated, decelerating phase 
of the past universe. 
} 
\label{aeta}
\end{center} 
\end{figure}

Figure~\ref{aeta} 
maps the conformal horizon.  When this curve has positive 
slope, the expansion is decelerating, $\ddot a<0$ and comoving 
wavelengths (e.g.~of density perturbations) enter the 
horizon.  More recently the curve has negative slope; 
this is the signature of accelerating expansion.  In terms of 
inflation we would say that wavelengths exit the horizon 
then (the horizon appears to shrink).  The dashed lines show the 
reconstruction possible by 
SNAP observing supernovae out to $z=1.7$.  SNAP clearly has 
the ability to probe not only the current accelerating epoch 
(``late time inflation'') but to reach back into the decelerating 
phase and to map the transition between them.

\section{Beyond Dark Energy}\label{sec.bey}

Mapping the physical time evolution of the scale factor relies on 
translating the observations into the Hubble 
parameter $H(z)$.  The Friedmann 
equation (\ref{hrho}) related this to the matter and dark 
energy densities. 
But ideally we would like to use the data to test the 
Friedmann equations of general relativity or alternate explanations 
for the acceleration besides dark energy.  Here we lay the 
framework for such options and give some examples. 

\subsection{Higher dimension theories}\label{sec.brane}

In the braneworld scenario of Deffayet et al. \cite{deff}, 
gravitation 
leaks from the 4-dimensional brane we experience out into the 
5-dimensional bulk.  The new expansion evolution 
equation is 
\beq 
H^2(z)=-ka^{-2}+\left[\sqrt{(8\pi/3)\rho+1/(2r_c)^2}+1/(2r_c)\right]^2,
\label{hbrane}\eeq 
where $r_c=M_{Pl}^2/(2M_5^3)$ is the crossover scale on which 
gravity leaks into the bulk, defined in terms of the usual Planck 
scale $M_{Pl}$ and the 5-dimensional Planck scale $M_5$.  It is 
convenient to introduce a dimensionless energy density $\omc=1/ 
(4H_0^2r_c^2)$.  
Flatness imposes $\omc=(1-\om)^2/4$. 

SNAP data will be able to constrain the parameter uncertainties to 
$\sigma(\om)=0.008$, or $\sigma(\omc)=0.003$ for a fiducial 
$\om=0.3$.  
This corresponds to a determination of the crossover scale 
of $r_c=1.43\pm0.015\,H_0^{-1}$.  If other observations, 
analyzed within this braneworld picture, disagreed with this 
narrow range then this theory could be ruled out.  Fig.~\ref{atmodsn} 
illustrates how the expansion history for a 
braneworld model differs from that for dark energy. 

Note that other higher dimension models discussed, such as 
the Randall-Sundrum type 2 model \cite{rs2,bine} and so-called 
Cardassian expansion \cite{freese}, reduce to the already 
discussed dark energy prescription for the recent expansion 
history.  

\subsection{Chaplygin gas}\label{sec.chap}

A very different possibility to explain the acceleration of the 
universe is the Chaplygin gas \cite{fabris}, proposed to unify 
dark energy and dark matter.  Its pressure $p\sim -1/\rho$ gives 
a solution that at early times behaves like nonrelativistic 
matter and at late times like a cosmological constant.  The 
expansion equation becomes 
\beq 
[H(z)/H_0]^2=\Omega_m(1+z)^3+(1-\Omega_m)\sqrt{A+(1-A)(1+z)^6}. 
\eeq 
The factor $A$ is interpreted as the sound velocity squared of the 
Chaplygin gas.  As $A\to1$, it reduces to the cosmological constant. 

SNAP data will be able to constrain the parameter uncertainties to 
$\sigma(\Omega_M)$$=+0.005,-0.017$ and $\sigma(A)=-0.005$ for a fiducial 
$\Omega_m=0.3$, $A=1$ (though the results are fairly insensitive to 
$A$).  Note 
dark matter in the form of ``quintessence clumps'' would require $A$ 
significantly smaller than unity.

\section{Conclusion} 

The geometry, dynamics, and composition of the universe are intertwined 
through the theory of gravitation governing the expansion of the universe. 
By precision mapping of the recent expansion history we can hope to 
learn about all of these.  The brightest hope for this in the near 
future is the next generation of distance-redshift measurements through 
Type Ia supernovae that will reach out to $z\approx1.7$.  

Just as the thermal history of the early universe taught us much about 
cosmology, astrophysics, and particle physics, so does the recent 
expansion history have the potential to greatly extend our physical 
understanding.  With the new parametrization of dark energy suggested 
here, one can study the effects of a time varying equation of state 
component back to the decoupling epoch of the cosmic microwave background 
radiation.  But even beyond dark energy, exploring the expansion history 
provides us cosmological information in a model independent way, allowing 
us to examine many new physical ideas.  From two numbers we have 
progressed to mapping the entire dynamical function $a(t)$, to the 
brink of a deeper understanding of the dynamics and fate of the universe. 

\begin{acknowledgments} 
This work was supported at LBL by the Director, Office of Science, 
DOE under DE-AC03-76SF00098.  I thank Alex Kim, Saul Perlmutter, and 
George Smoot 
for stimulating discussions and Ramon Miquel and Nick Mostek for 
stimulating computations. 
\end{acknowledgments}


\begin{thebibliography}{99}

\bibitem{san61}
	A.~Sandage, Ap.J.~{\bf 133}, 355 (1961) 

\bibitem{san88}
	A.~Sandage, Ann.~Rev.~A\&A {\bf 26}, 561 (1988) 

\bibitem{lin89} 
	E.V.~Linder, A\&A {\bf 206}, 175 (1988) 

\bibitem{lin97}
	E.V.~Linder, {\it First Principles of Cosmology} 
	(Addison-Wesley, London, 1997)  

\bibitem{teg}
	M.~Tegmark, Science {\bf 296}, 1427 (2002), astro-ph/0207199

\bibitem{kt}
	E.W.~Kolb and M.S.~Turner, {\it The Early Universe} 
	(Addison-Wesley, New York, 1990) 

\bibitem{astier}
	P.~Astier, astro-ph/0008306

\bibitem{ht}
	D.~Huterer and M.S.~Turner, Phys.~Rev.~D {\bf 64}, 123527 (2001), 
	astro-ph/0012510 

\bibitem{braxm}
	P. Brax and J. Martin, Phys. Lett. B {\bf 468}, 40 (1999), 
	astro-ph/9905040

\bibitem{snap}
	SNAP (http://snap.lbl.gov) parameter error estimates are for 
2000 SNe between $z=0.1-1.7$ and 300 SNe at $z<0.1$ from the Nearby 
Supernova Factory (http://snfactory.lbl.gov), plus a prior on 
$\Omega_M$ of 0.03.  Some numerical results were kindly provided by 
R.~Miquel and N.~Mostek; see Miquel, Mostek, \& Linder 2002, in 
preparation, for a description of the Monte Carlo fitter. 

\bibitem{planck}
	Planck: http://astro.estec.esa.nl/Planck 

\bibitem{fhlt} 
	J.A.~Frieman, D.~Huterer, E.V.~Linder, and M.S.~Turner, 
	submitted to Phys.~Rev.~D, astro-ph/0208100 

\bibitem{welal} 
	J.~Weller and A.~Albrecht, Phys.~Rev.~D {\bf 65}, 103512 (2002), 
	astro-ph/0106079

\bibitem{deff} 
	C.~Deffayet, G.~Dvali, G.~Gabadadze, Phys.~Rev.~D {\bf 65}, 
	044023 (2002), astro-ph/0105068

\bibitem{rs2}
	L.~Randall and R.~Sundrum, Phys.~Lett.~B {\bf 83}, 4690 (1999), 
	hep-th/9906064

\bibitem{bine}
	P.~Binetruy, C.~Deffayet, U.~Ellwanger, D.~Langlois, 
	Phys.~Lett.~B {\bf 477}, 285 (2000), hep-th/9910219

\bibitem{freese} 
	K.~Freese and M.~Lewis, astro-ph/0201229 

\bibitem{fabris}
	J.C.~Fabris, S.V.B.~Gon\c{c}alves, and P.E.~de Souza, 
	astro-ph/0207430 

\end{thebibliography}
\end{document}